\documentclass[pdflatex,sn-mathphys]{sn-jnl}

\jyear{2022}%

\theoremstyle{thmstyleone}%
\theoremstyle{thmstyletwo}%

\theoremstyle{thmstylethree}%

\begin{document}

\title[Hybrid Deep Feature-Based Image Registration]{A Hybrid Deep Feature-Based Deformable Image Registration Method for Pathological Images}


\author[1,3]{\fnm{Chulong} \sur{Zhang}}
\author[2]{\fnm{Yuming} \sur{Jiang}}
\author[4]{\fnm{Na} \sur{Li}}
\author[5]{\fnm{Zhicheng} \sur{Zhang}}
\author[2]{\fnm{Md Tauhidul} \sur{Islam}}
\author[1]{\fnm{Jingjing} \sur{Dai}}
\author[1]{\fnm{Lin} \sur{Liu}}
\author[1]{\fnm{Wenfeng} \sur{He}}
\author[1]{\fnm{Wenjian} \sur{Qin}}
\author[1]{\fnm{Jing} \sur{Xiong}}
\author[1]{\fnm{Yaoqin} \sur{Xie}}
\author*[1]{\fnm{Xiaokun} \sur{Liang}}\email{xk.liang@qq.com / xk.liang@siat.ac.cn}

\affil*[1]{\orgdiv{Shenzhen Institute of Advanced Technology}, \orgname{Chinese Academy of Sciences}, \orgaddress{\city{Shenzhen}, \postcode{518055}, \state{Guangdong}, \country{China}}}

\affil[2]{\orgdiv{Department of Radiation Oncology}, \orgname{Stanford University}, \orgaddress{\city{Stanford}, \postcode{94305}, \state{CA}, \country{USA}}}

\affil[3]{\orgdiv{School of Mathematics}, \orgname{Sun Yat-sen University}, \orgaddress{\city{Guangzhou}, \postcode{510275}, \state{Guangdong}, \country{China}}}

\affil[4]{\orgdiv{Department of Biomedical Engineering}, \orgname{Guangdong Medical University}, \orgaddress{\city{Dongguan}, \postcode{523808}, \state{Guangdong}, \country{China}}}

\affil[5]{\orgname{Xiaohe Healthcare ByteDance}, \orgaddress{\city{Guangzhou}, \postcode{510000}, \state{Guangdong}, \country{China}}}


\abstract{Pathologists need to combine information from differently stained pathology slices for accurate diagnosis. Deformable image registration is a necessary technique for fusing multi-modal pathology slices. This paper proposes a hybrid deep feature-based deformable image registration framework for stained pathology samples. We first extract dense feature points via the detector-based and detector-free deep learning feature networks and perform points matching. Then, to further reduce false matches, an outlier detection method combining the isolation forest statistical model and the local affine correction model is proposed. Finally, the interpolation method generates the deformable vector field for pathology image registration based on the above matching points. We evaluate our method on the dataset of the Non-rigid Histology Image Registration (ANHIR) challenge, which is co-organized with the IEEE ISBI 2019 conference. Our technique outperforms the traditional approaches by 17$\%$ with the Average-Average registration target error (rTRE) reaching 0.0034. The proposed method achieved state-of-the-art performance and ranked 1st in evaluating the test dataset. The proposed hybrid deep feature-based registration method can potentially become a reliable method for pathology image registration.}




\maketitle

\section{Introduction}\label{sec1}
\begin{figure}
\centerline{\includegraphics[width=1\columnwidth]{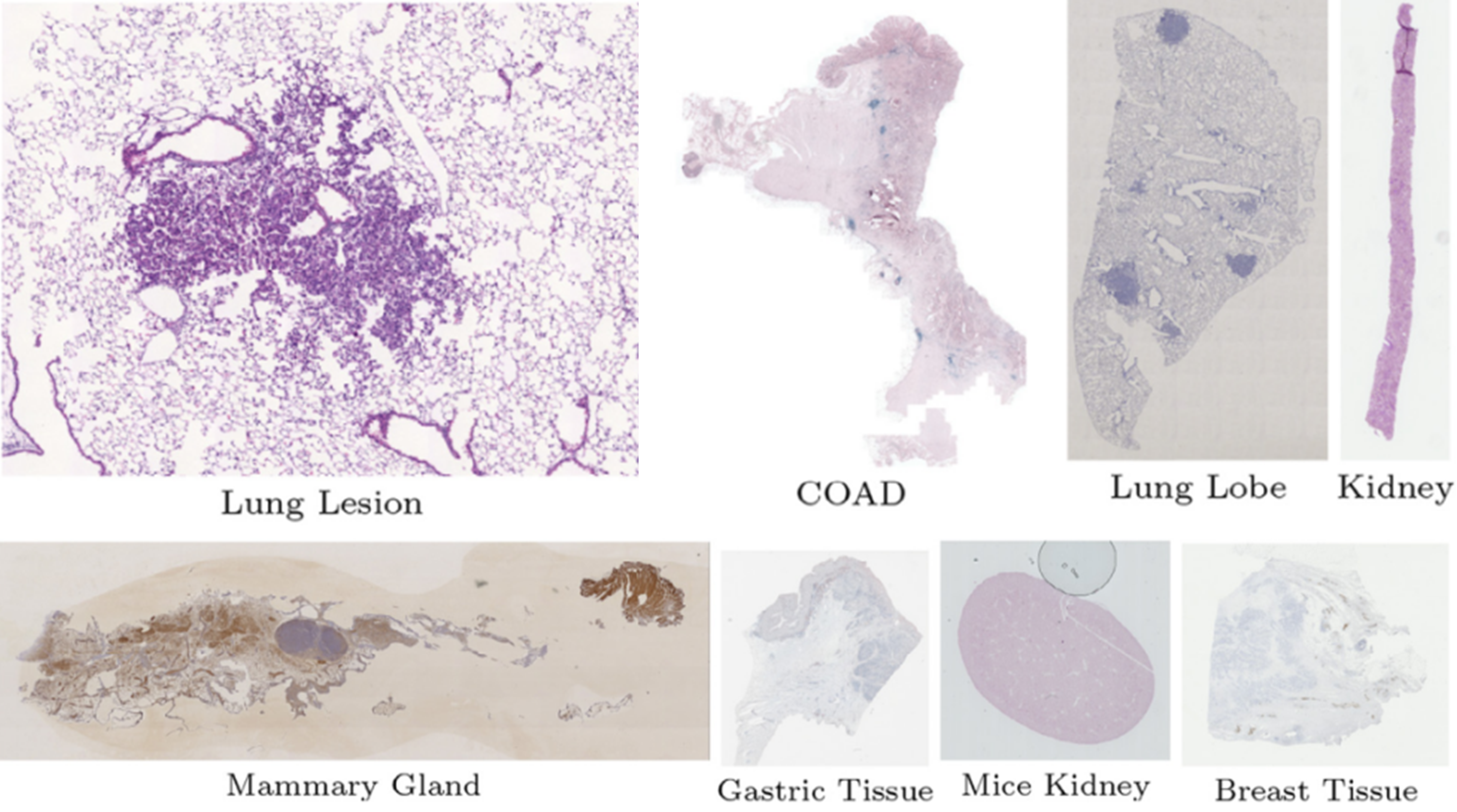}}
\caption{\centering{ Examples of pathology images of each site.}}
\label{fig:sample}
\end{figure}

Registration of pathology images is vital for current histology image analysis and decision-making. In clinics, the registration of pathology images is commonly performed based on the three-dimension (3D) reconstruction of scanned two-dimension (2D) thin slices, creating a high-resolution pathology image. The fusing technique is then performed on separately stained slices \cite{b4}. The challenge of this process is that pathology images are characterized by high resolution, large deformation, repetition, and complex textures \cite{b1}. Examples of some pathology images of different sites are shown in Fig. \ref{fig:sample}

The conventional methods used for medical image registration are not suitable for pathology images \cite{b19,b20,b21,b22,b23} due to the properties discussed earlier. Traditional medical image registration techniques include intensity-based, feature-based, and segmentation-based methods. In recent years, researchers have developed a number of non-rigid registration methods, such as bUnwarpJ \cite{b19}, NiftyReg \cite{b20}, RVSS \cite{b19}, ANTs \cite{b21}, DROP \cite{b22}, and Elastix \cite{b23}. The above methods were tested in the Automated Non-Rigid Histological Image Registration (ANHIR) challenge dataset \cite{b1,b2,b3}. However, the results obtained were unsatisfactory.

\begin{figure*}
\centerline{\includegraphics[width=1\columnwidth]{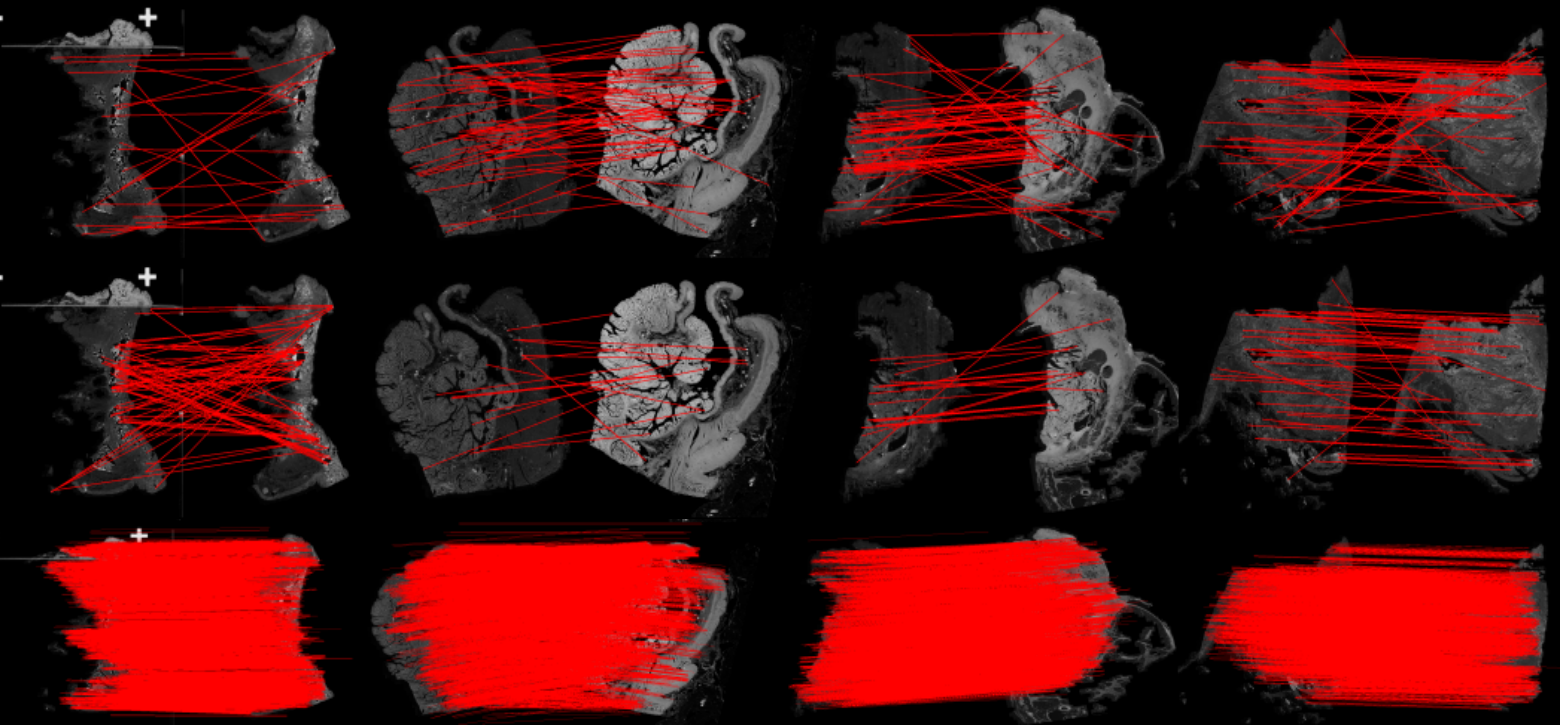}}
\caption{\centering{ A visual comparison of the results of our method with those of traditional manual algorithms.}}
\label{fig:compare}
\end{figure*}

Fraunhofer MEVIS was the method with the best performance on the ANHIR dataset \cite{b1,b7}. The authors provide a non-rigid registration framework using normalized gradient fields as a similarity measure. This framework includes the steps of initial rotation search, iterative affine, and b-spline-based registration. A deep learning technique was also proposed for analyzing the ANHIR dataset \cite{b1,b8}. The researchers optimized the network using manually defined signposts after re-sampling the images to a relatively low resolution. Although the technique is faster than other competing methods, the accuracy of this method is not convincing. Subsequently, an unsupervised medical image registration network based on deep learning was proposed \cite{b4,b5,b6,b34,b35,gan2022probabilistic}. The framework is separated into three sections: initial alignment, affine alignment, and non-rigid registration. They employ global normalized cross-correlation (NCC) for initial alignment and affine registration and patch-based normalized cross-correlation for non-rigid registration. The smooth of the deformable vector field (DVF) is used as the regularization term for the non-rigid registration. Although the approach significantly improved registration speed, it failed to deliver high accuracy.

In intensity-based registration, iterative search strategies or neural networks are employed to optimize similarity metrics, including mean square error (MSE) and NCC, et al. As a result, the DVF corresponding to the fixed and moving images are obtained. Another alternative is the feature-based registration technique. Its implementation includes the following steps: firstly, feature points of the image are obtained, and feature descriptors are used to describe the feature points; then, matching is performed according to the descriptors; finally, transform fields are generated by interpolation after point matching. The main drawback of learning-based registration methods depending on similarity measures is that they still inherit the iterative nature of traditional deformable image registration (DIR) frameworks and share identical or similar objective functions. However, traditional methods can minimize this objective function for each case separately \cite{b4}. Therefore, in order to achieve higher accuracy in the registration, especially in tasks with large shape variations, we need to give the algorithm the ability to bypass the risk of getting trapped in a locally optimal solution. The feature-based method is one of the solutions.

Feature-based methods work in both manual extraction, and deep learning-based registration approaches \cite{b16}. Traditional methods such as SIFT \cite{b9}, ORB \cite{b10} use gradient, grayscale, and other information to manually design feature points for matching. Nevertheless, this technique concentrates on local features and yields sparse feature points. In contrast, deep learning-based feature extraction algorithms consider characteristics of diverse sizes and scales, contributing to more comprehensive features. There are two types of deep learning techniques: detector-based and detector-free approaches \cite{b11}. Whether a feature extraction network is required determines how these two categories vary from one another. The detector-based approach uses the response values of the network to detect interest points. It constructs descriptors using features extracted by a convolutional neural network (CNN) and U-Net, such as SuperPoint \cite{b12}, DISK \cite{b13}. These interest points and descriptors then match and get the corresponding feature pairs. Detector-free techniques, such as correspondence transformer-based image matching network \cite{b14}, and LOFTR \cite{b11}, employ the designed network for end-to-end matching without requiring a dedicated detector to identify interest sites. The detector-based approach focuses more on the interest points detected by the detector. The detector-free approach, on the other hand, focuses more on global information because it performs end-to-end training. In this paper, the two feature methods are combined to obtain different point pairs.

However, pre-matching or affine transformation of medical images based on traditional hand-crafted feature points (SIFT, SURF \cite{b27}, ORB) is common. Recently, some work has proven that feature-based approaches, particularly those based on deep learning, perform well for cross-modality medical image matching \cite{b17}. The study states the results of several feature matching methods for typical multi-modal images in medical research fields, such as MRI-PET, CT-SPECT, etc. Additionally, it demonstrates that implementing deep feature approaches for medical image registration is viable and beneficial.

Two main issues limit the accuracy of conventional feature-based approaches for registration: 1) feature points and matched point pairs are excessively sparse, resulting in unacceptable errors between the interpolated estimated DVF and the real DVF. 2) there might be substantial mismatches, which severely diminish the accuracy of the DVF. We propose a hybrid deep learning feature-based registration framework to tackle these challenges. We integrate detector-based and detector-free networks to extract features and sufficiently match dense pairs of feature points. In addition, we construct an outlier detection module for error matching pair detection based on the local and overall consistency of medical images.The Fig. \ref{fig:compare} shows a qualitative comparison of the matching point pairs obtained by our method with the results obtained by traditional manual features. From top to bottom, SIFT, ORB and our method are shown in order.

We validate our approach on the ANHIR dataset. We observed that employing model parameters pre-trained for natural image datasets during inference achieved fairly good performance. Furthermore, we combine the robust results of two separate networks, the outputs after anomaly detection as pseudo-labels, and the landmarks supplied in the dataset as ground truth to co-construct the training set to fine-tune the network. These further enhance the network performance.
We have three major contributions.

\begin{itemize}
\item To solve the problem of sparse and erroneous matching points in traditional feature-based medical image registration, we proposed a hybrid deep feature-based image registration method for pathology image registration. The proposed method achieved state-of-the-art performance.
\item In our experiments, we found that the feature matching network training based on natural images can be well migrated to the scenes of medical images and obtained quite a high accuracy. The results inspired us to use the large dataset in natural images to solve the tricky problems in medical images.
\item We converted the erroneous matching problem into a global and local outlier detection problem. An outlier detection method was proposed combining the isolation forest statistical model and the local affine correction model.
\end{itemize}

\section{Method}

\begin{figure*}
\centerline{\includegraphics[width=1\columnwidth]{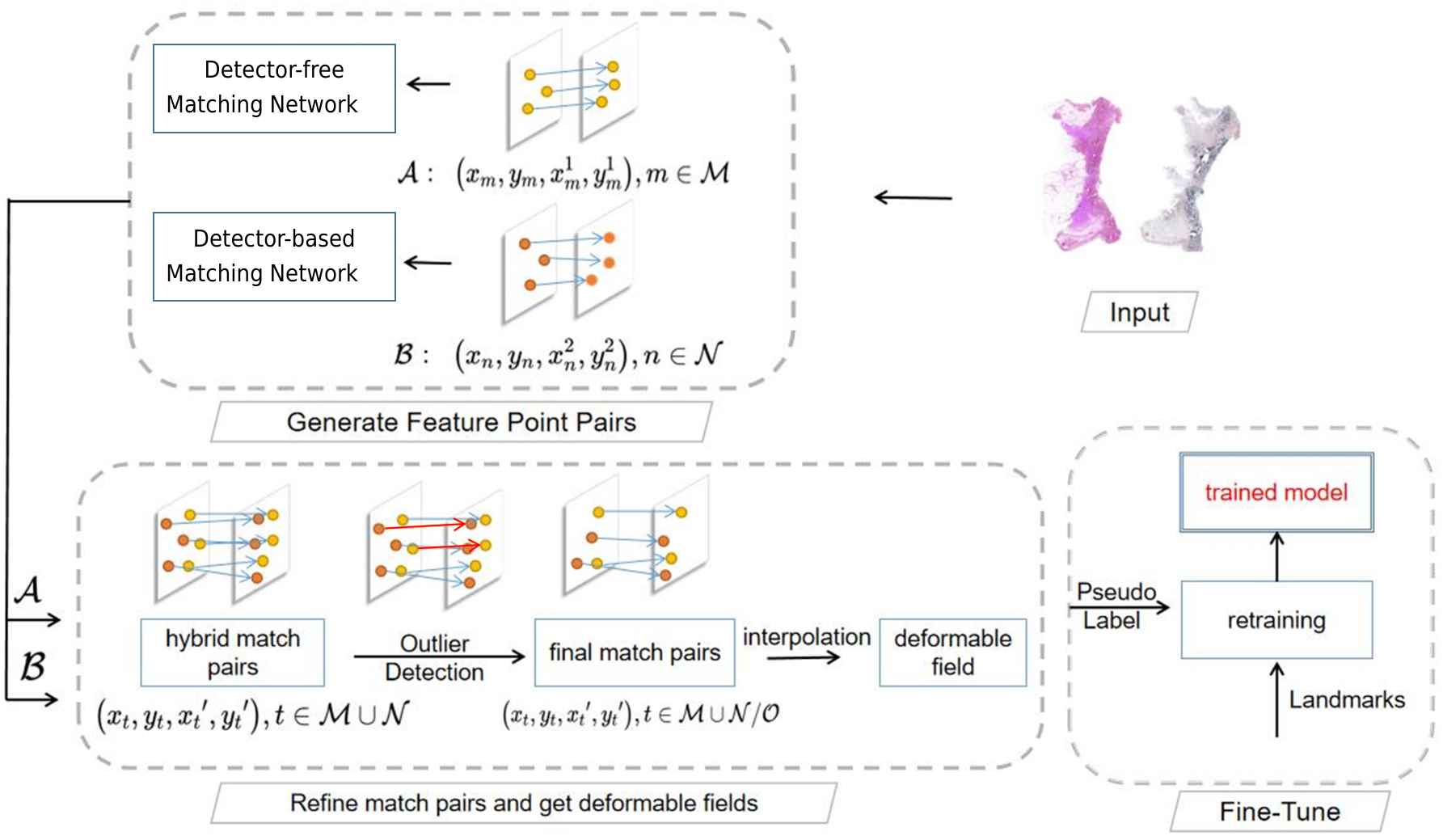}}
\caption{Workflow of the proposed method. Generate Feature Point Paris: We extract multiple pairs of feature points from different networks. Refine match pairs and get deformable fields: The point pairs are combined and passed into the outlier detection module. Finally, new point pairs are obtained, and the DVF is generated by interpolation. Fine-Tune: First, we employ a pre-trained model in natural images on our inference. Then, we use the inference output to construct pseudo-labels and landmarks to train the network together. The final trained model is obtained to achieve more accurate results.}
\label{fig:workflow}
\end{figure*}

We provide a framework for hybrid deep feature-based registration. Our framework is separated into feature point pair generation, anomaly matching detection, and DVF generation. In our initial inference procedure, we employ pre-trained models that have been learned directly on the natural image dataset. The outputs of the model and the landmarks in the training set are combined with fine-tuning the model, yielding the newly trained model. The new model provides more accurate results. Fig. \ref{fig:workflow} depicts the detailed flow chart. We extract various feature point pairs in the feature point pairs generation module. Feature point pair A: $(x_m,y_m,x_m^1,y_m^1)$,$m \in\mathcal{M}$ and feature point pair B: $(x_n,y_n,x_n^2,y_n^2 )$ ,$n \in \mathcal{N}$, $(x_m,y_m)$ and $(x_n,y_n)$ are the coordinates of the feature points in the source image, $(x_m^1,y_m^1)$ and $(x_n^2,y_n^2)$ are the coordinates of the feature point pairs in the predicted target image.

\begin{figure}
\centerline{\includegraphics[width=\columnwidth]{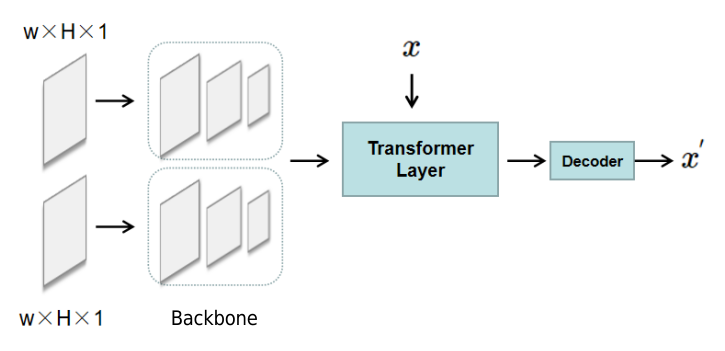}}
\caption{The Detector-Free Matching Network comprise the feature extraction and Transformer module. After approaching the CNN network to extract features, the two images go through an encoding-decoding-decoding procedure to get the coordinates x in the fixed image corresponding to ${x}^{'}$ in the moving image.}
\label{fig:cotr}
\end{figure}

We integrate the feature point pairs produced by the two networks to create new point pairs $(x_t,y_t,x_t^{'},y_t{'}), t \in \mathcal{M}\cup \mathcal{N}$. An outlier diagnosis module is proposed in anomaly matching detection. The DVF generation module combines an isolated forest model and a local affine correction model. The local affine correction model is based on global and local image consistency principles. The anomalous matches are eliminated while using this strategy. We employ the final obtained point pairs as pseudo-labels to fine-tune the original model together with the ground-truth point pairs provided by the dataset. As a result, the final obtained model performs a more accurate inference.

\subsection{Detector-Free Matching Network}
First, the two input images are resized to 256 × 256. Subsequently, the two are combined side-by-side using the same CNN backbone extracted to dimension 16 × 16 × 256 feature map $\mathcal{E}$. Subsequently, the coordinate function is $\Omega$ encoded via the position encoder $\mathcal{P}$, and the above two are combined to obtain the contextual feature map $c \in \mathcal{R}_{16*32*256}$. The network architecture of Detector-Free Matching Network is shown in Fig. \ref{fig:cotr}\cite{b14}

Then $c$ is fed to the transformer encoder $\mathcal{T_{\mathcal{E}}}$, and the position-encoded $\boldsymbol{x}$ is decoded by the transformer decoder $\mathcal{T_{\mathcal{D}}}$. Finally, the network output $\boldsymbol{x}^{'}$ is obtained after an MLP, and the above process can be described in the following form.

\begin{footnotesize} 
 	\begin{equation}
 		\begin{aligned}
 			\boldsymbol{x}^{\prime}=\mathcal{F}_{\boldsymbol{\Phi}}\left(\boldsymbol{x} \mid \boldsymbol{I}, \boldsymbol{I}^{\prime}\right)=\mathcal{D}\left(\mathcal{T}_{\mathcal{D}}\left(\mathcal{P}(\boldsymbol{x}), \mathcal{T}_{\mathcal{E}}(\mathbf{c})\right)\right)
 		\end{aligned}
 	\end{equation}
 \end{footnotesize}

This network can obtain highly accurate matching relationships by iterating using ${\mathcal F_{\Phi}}$. The initial value of the next iteration is a scaled-up version of the previous prediction after cropping, scaling the image down to 256 × 256. This way, we can achieve matching at any scale and fully use the information at each scale.

The goal of the network is to minimize the error in the matching as well as the mutual consistency error of the matches. 

\begin{footnotesize} 
 	\begin{equation}
 		\begin{aligned}
 			\underset{\boldsymbol{\Phi}}{\arg \min } \underset{\left(\boldsymbol{x}, \boldsymbol{x}^{\prime}, \boldsymbol{I}, \boldsymbol{I}^{\prime}\right) \sim \mathcal{D}}{\mathbb{E}} \mathcal{L}_{\text {corr }}+\mathcal{L}_{\text {cycle }}
 		\end{aligned}
 	\end{equation}
 \end{footnotesize}
 
\begin{footnotesize} 
 	\begin{equation}
 		\begin{aligned}
 			\mathcal{L}_{\text {corr }}=\left\|\boldsymbol{x}^{\prime}-\mathcal{F}_{\boldsymbol{\Phi}}\left(\boldsymbol{x} \mid \boldsymbol{I}, \boldsymbol{I}^{\prime}\right)\right\|_{2}^{2}
 		\end{aligned}
 	\end{equation}
 \end{footnotesize}
 
\begin{footnotesize} 
 	\begin{equation}
 		\begin{aligned}
 			\mathcal{L}_{\text {cycle }}=\left\|\boldsymbol{x}-\mathcal{F}_{\boldsymbol{\Phi}}\left(\mathcal{F}_{\boldsymbol{\Phi}}\left(\boldsymbol{x} \mid \boldsymbol{I}, \boldsymbol{I}^{\prime}\right) \mid \boldsymbol{I}, \boldsymbol{I}^{\prime}\right)\right\|_{2}^{2}
 		\end{aligned}
 	\end{equation}
 \end{footnotesize}

Where $\mathcal{D}$ denotes the training set with true matching relationships;  $\mathcal{L}_{\text {corr}}$ denotes the error in matching estimates, and $\mathcal{L}_{\text{cycle}}$ denotes the mutual consistency error.

The network needs to be pre-trained on the natural image dataset MegaDepth \cite{b30}. The dataset contains 200 models of landmarks worldwide with dense 3D reconstruction, while 150,000 reconstructed images are presented. After filtering, 130,000 usable images are retained; out of these 130,000 images, about 100,000 are absolute depth data, and 30,000 are relative depth data. We can get the one-to-one correspondence of the position relationship of the image pairs in the dataset for training.

\subsection{Detector-based Matching Network}

The detector-based matching network comprises two networks: a feature point detecotr network based CNN \cite{b12} and a feature point matching network based graph network \cite{b15}. 

\begin{figure}
\centerline{\includegraphics[width=\columnwidth]{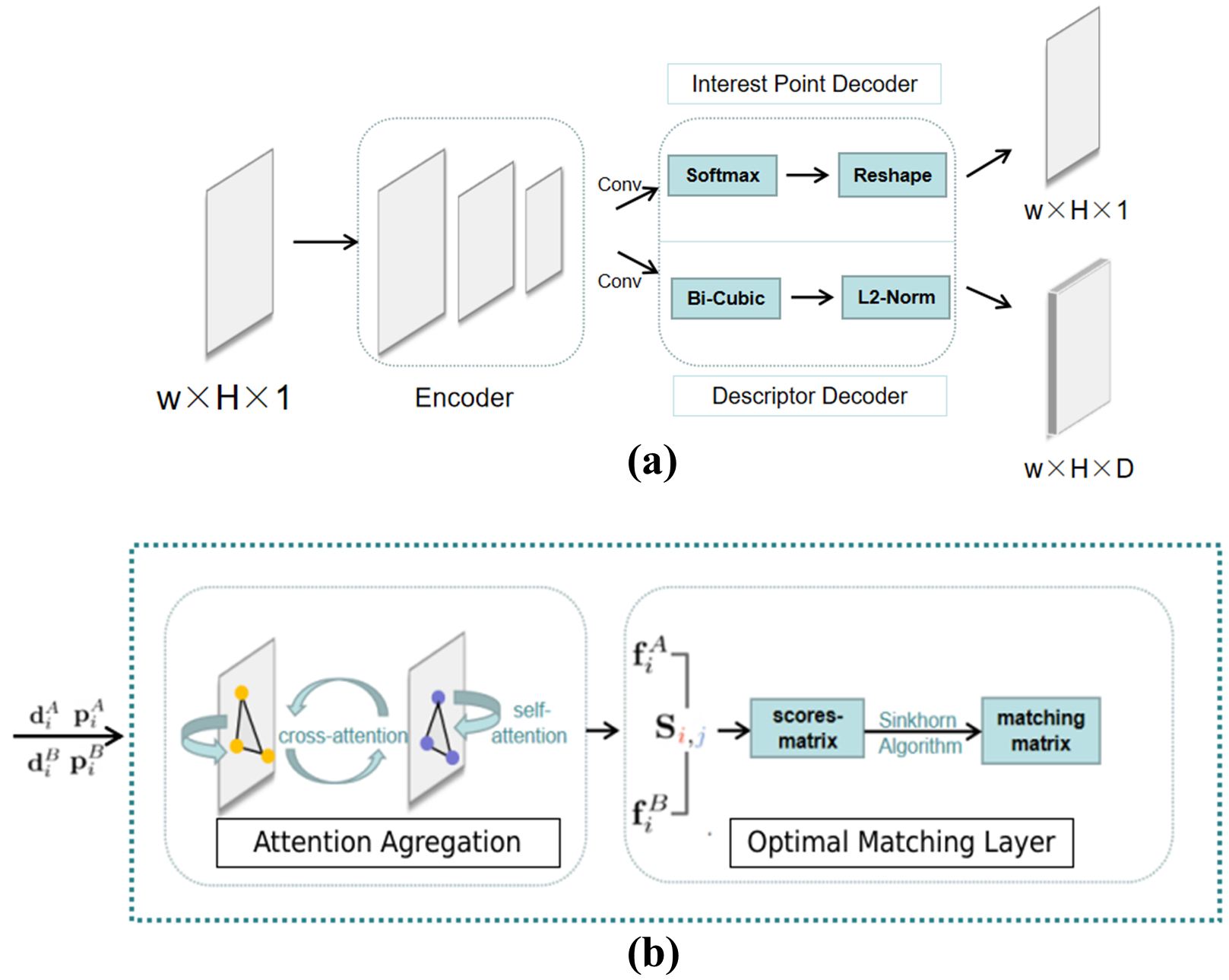}}
\caption{Two parts of detector-based matching network. The first part is the detector network (a), a self-supervised feature point extraction and descriptor construction network. The second part is the matching network (b), a feature matching network based on graph networks.}
\label{fig:sur}
\end{figure}

The first part of the network is a deep feature point extraction and descriptor generation network. The structure is shown in Fig. \ref{fig:sur}. A feature point detection network is a decoder. It determines the probability of each pixel in an image being a feature point by estimating its likelihood. A decoder is also a component of the descriptor detection network. First, a semi-dense descriptor is learned, then a bi-trivial interpolation is used to acquire the entire descriptor. Finally, L2-normalizes are performed to generate a unit-length descriptor. The entire loss function is presented as follows:

\begin{footnotesize} 
 	\begin{equation}
 		\begin{aligned}
 			\mathcal{L}\left(X, X^{\prime} ; Y^{\prime} ; \quad D, D^{\prime} ; S\right)=\mathcal{L}_{p}(X, Y)+\mathcal{L}_{d}\left(X^{\prime}, Y^{\prime}\right)+\lambda \mathcal{L}_{d}\left(D, D^{\prime}, S\right)
 		\end{aligned}
 	\end{equation}
 \end{footnotesize}

Where $\mathcal{L}_{p}$ is the loss function for the feature points, and $\mathcal{L}_{d}$ is the loss function for the descriptors, with coefficients to balance the two weights.

The matching network is a feature matching network based on Graph Neural Networks and attention mechanisms. Our high-dimensional perceptron performs location encoding by encoding the location information $\mathbf{p}_i=(x,y,c)$ into the descriptor $\mathbf{d}$. Then, feeding the encoded descriptor into a graph network based on self-attention and cross-attention mechanisms after two attention layers: (i) self-attention, which enhances the receptivity of local descriptors; and (ii) cross-attention, which enables cross-image information flow. Finally, the Sinkhorn Algorithm is used to obtain the optimal matching matrix.

The detector network is trained with generated pseudo-ground truth data and image pairs obtained by deforming many images on the MS-COCO \cite{b28}  dataset. The matching network uses many indoor paired images provided by the ScanNet \cite{b29} dataset as the training set. 

\subsection{Outlier Detection Model}

Outlier detection is proposed to tackle the problem of detecting the wrong point pair. For example, we commonly employ random sampling consensus \cite{b31,b32} to eliminate outliers in image matching. On the other hand, random sampling consensus is based on the simple premise that all points in an image meet the same Homography or affine transformation, which is incorrect in the case of sophisticated histology image deformations.

\begin{figure}
\centerline{\includegraphics[width=\columnwidth]{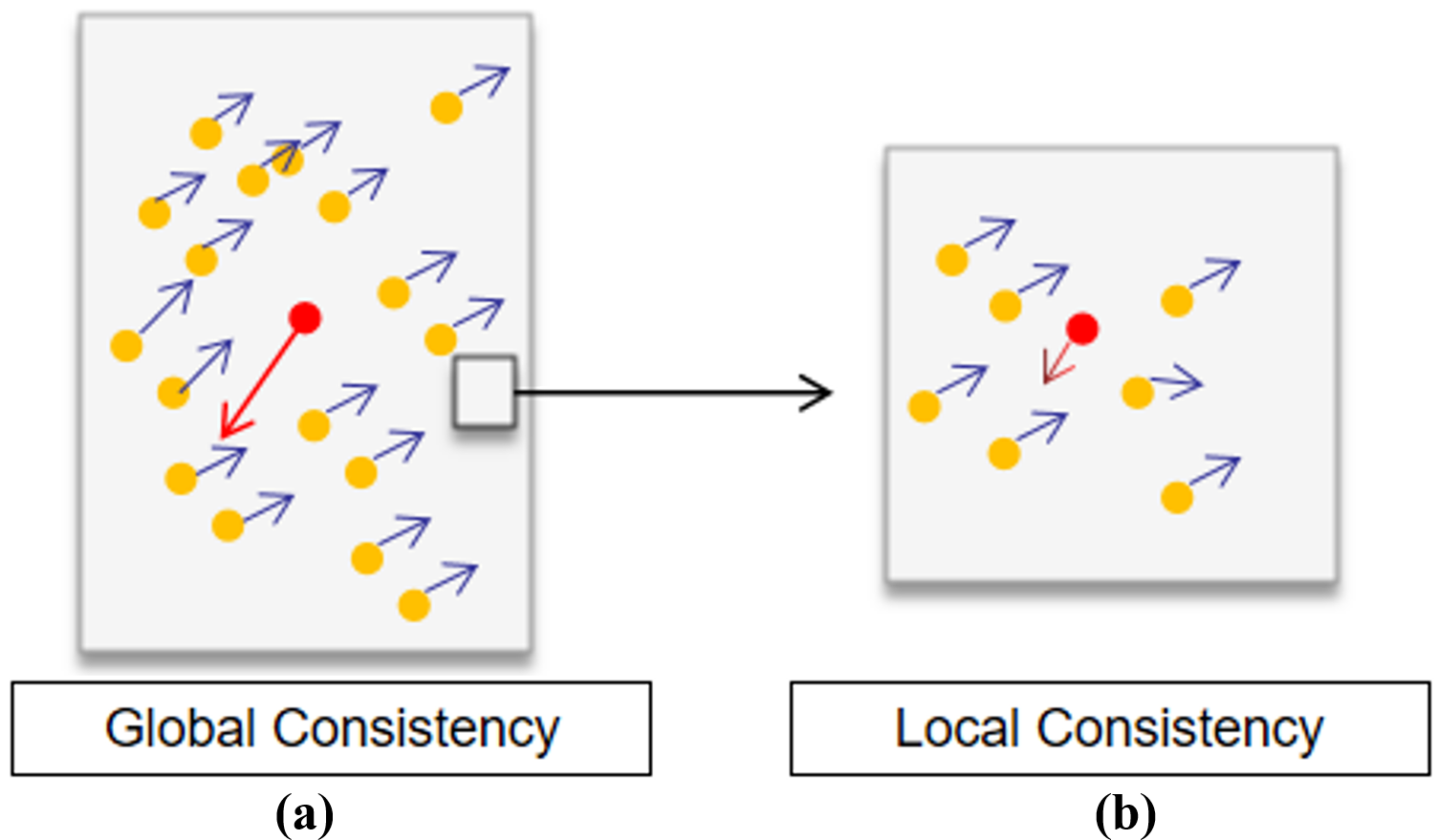}}
\caption{The following two cases are considered as the displacement of feature points is anomalous, 1) the displacement of feature points is too different from the global displacement, and 2) the displacement of feature points is too different from the displacement of other feature points of the local image. (a) represents global consistency and (b) represents local consistency.}
\label{fig:detection}
\end{figure}

We then assume the following hypotheses for point displacements in pathology images: 1) The overall displacement of points in the whole image is consistent, i.e., the direction and distance of the displacements of all points in the image will not differ too much, which may be demonstrated by the fact that a simple affine transform estimate is adequate to reach high accuracy. 2) The local distortion of the image will not vary too much, i.e., the displacement of a point and the displacement near that point should be consistent. The overall consistency and local consistency are visibly illustrated in Fig. \ref{fig:detection} Based on the above, we build an Isolation Forest statistical model \cite{b18,b33} and a Local affine correction model for outlier detection. We use the Isolation Forest model to solve the global consistency problem. We employ  $s_{i}=(\Delta x,\Delta y)$ to signify the displacement of the predicted coordinates from the initial coordinates.

\begin{footnotesize} 
 	\begin{equation}
 		\begin{aligned}
 			\Delta x=x_t^{'} - x_t,\Delta y=y_t{'} - y_t
 		\end{aligned}
 	\end{equation}
 \end{footnotesize}

Then we obtain the set of displacements of the whole image feature points as

\begin{footnotesize} 
 	\begin{equation}
 		\begin{aligned}
 			S=\left\{s_{1}, \cdots, s_{n}\right\}
 		\end{aligned}
 	\end{equation}
 \end{footnotesize}

According to our hypothesis, the erroneously matched points represent set S's outliers. We randomly choose the hyperplane and recursively split the data set S until any of the following requirements are fulfilled. (1) the tree achieves the maximum height; (2) there is only one sample on the node; (3) all features of the samples on the node are the same. The relationship between the height limit $l$ of the tree and the number of sub-samples $\psi$ is $l = \text { ceiling }\left(\log _{2}(\psi)\right)$.

The iForest model consists of two steps: the training phase, which forms isolated trees from sub-samples of the training set, and the testing phase, which computes anomaly scores for each test sample using isolated trees.

In the training phase, the iTree is built by recursively separating the training set until all samples are isolated or the tree has reached the given height. 

Then the empirical value of the number of sub-samples $\psi$ tree with the number limit $t$ can be obtained. From there, isolated samples are obtained by calculating the anomaly score of the test sample by the expected path length $E(h(x))$. $h(x)$ is the number of edges passed from the root node to the leaf nodes of the iTree. Finally, we utilize the average path length $c(n)$ of the tree and $E(h(x))$ to generate the anomaly scores of the samples $s(x, n)=2^{-\frac{E(h(x)}{c(n)}}$, from which we extract the anomaly values.

The specific algorithm is shown in Algorithm 1.

We filter out those sample points significantly different from the global displacement with the above algorithm. Nonetheless, the local anomaly problem in Hypothesis 2 has not been addressed. Therefore, we propose a local affine correction model to conquer this challenge.

\begin{figure}
\centerline{\includegraphics[width=\columnwidth]{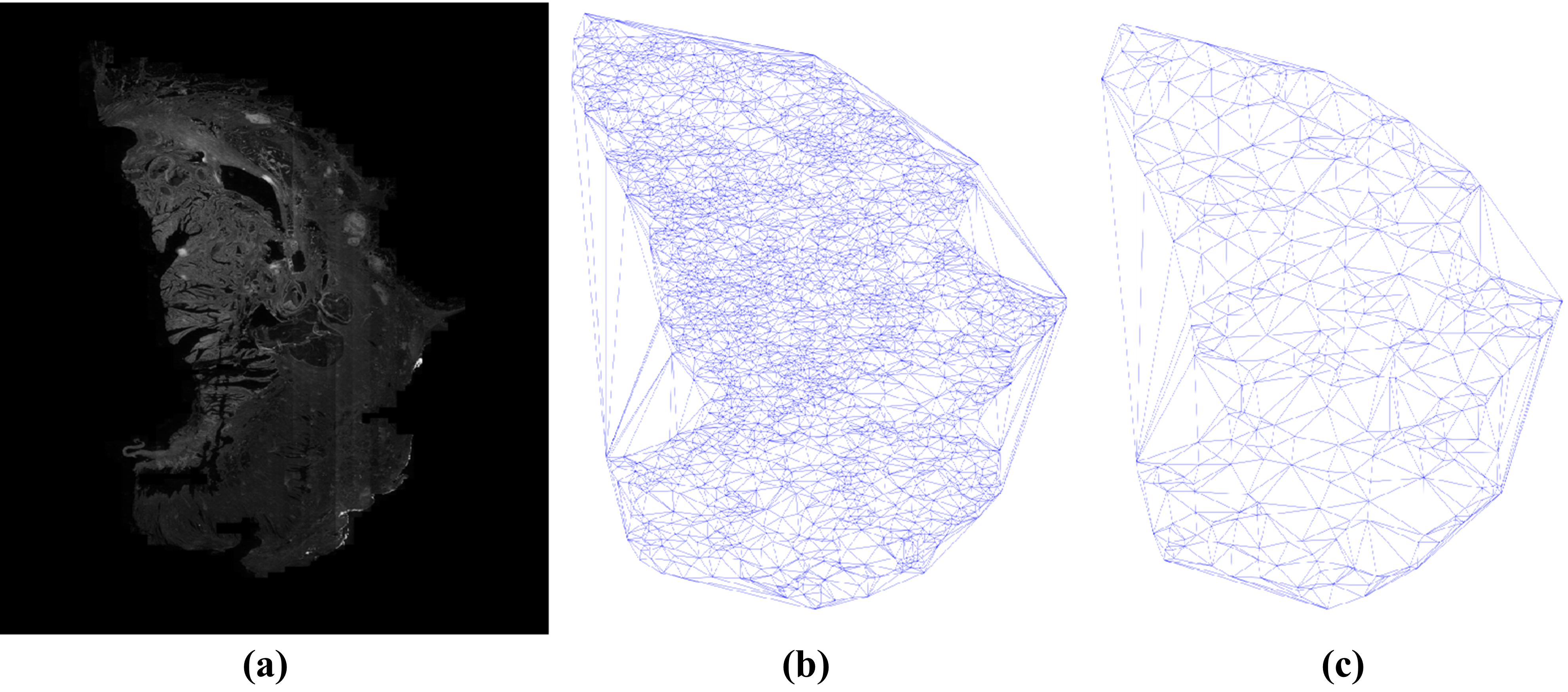}}
\caption{Triangle dissection results before and after sampling the feature points on the pathology image, where (b) and (c) represent the triangular dissection results before and after down-sampling of the original image (a). We formed triangles that spread over the pathology image after down-sampling. Each triangle may correspond to an affine transformation, and the points covered in the triangle should match the affine transformation of that triangle. }
\label{fig:affinedetect}
\end{figure}

First, we sample the feature points in the image randomly and evenly. Then, the sampled point set is constructed as a triangular dissection to lay out the triangular shape of the pathology image. The schematic diagram is shown in Fig. \ref{fig:affinedetect}. The competence of each triangle to construct an affine transformation depending on the displacement of feature points is readily ascertained. Therefore, the points covered by the triangles should mainly fulfill the affine transform if these transformations of points are valid.

$$
\begin{array}{l}
\hline \text { Algorithm 1: PathLength }(s, T, e) \\
\hline \text { Inputs : } s \text { - an instance, } T \text {-an iTree, } e \text { - current path length; } \\
\text { to be initialized to zero when first called } \\
\text { Output: path length of } s \\
\text { 1: if } T \text { is an external node then } \\
\text { 2: return } e+c(T \text {.size })\{c(.) \text { is defined in Equation 1 }\} \\
\text { 3: end if } \\
\text { 4: } a \leftarrow T . s p l i t A t t \\
\text { 5: if } s_{a}<T . \text { splitValue then } \\
\text { 6: return PathLength }(s, T \text {.left, } e+1) \\
\text { 7: else }\left\{s_{a} \geq \text { T.splitValue }\right\} \\
\text { 8: return PathLength }(s, T \text {.right }, e+1) \\
\text { 9: end if }\\
\hline
\end{array}
$$

We allow the point pair formed by some pair and the corresponding point be $(x_t,y_t,x_t^{'},y_t^{'})$, and the point pair obtained by the affine transformation covering that point at the ith sampling is $(x_{t},y_{t},\hat{x}^{'}_{ti},\hat{y}^{'}_{ti})$, then we note that 

\begin{footnotesize} 
 	\begin{equation}
 		\begin{aligned}
           s_t = \Sigma^{n}_{i = 0}\sqrt{(x_t^{'}-\hat{x}^{'}_{ti})^2+(y_t^{'}-\hat{y}^{'}_{ti})^2}
 		\end{aligned}
 	\end{equation}
 \end{footnotesize}

$s_t$ is the anomaly score of a point after n times of sampling. We discard the points with high anomaly scores after multiple sampling to obtain the matching points that satisfy local consistency.

\begin{figure}
\centerline{\includegraphics[scale=0.5]{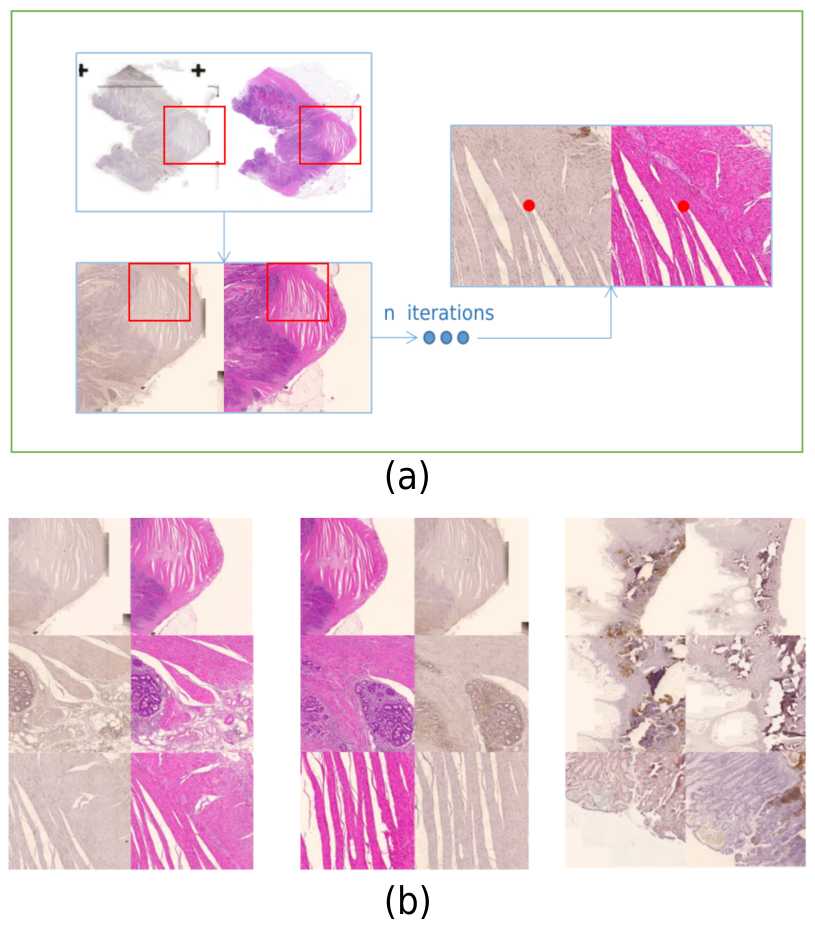}}
\caption{The image was down-sampled and cropped centered on the inferred points to get a new image for iterative inference. The above figure (a) depicts the iterative inference process. While the following figure (b) illustrates the pairs of patches constructed according to the iterative inference process.}
\label{fig:trainset}
\end{figure}

\begin{figure}
\centering
\includegraphics[scale=0.16]{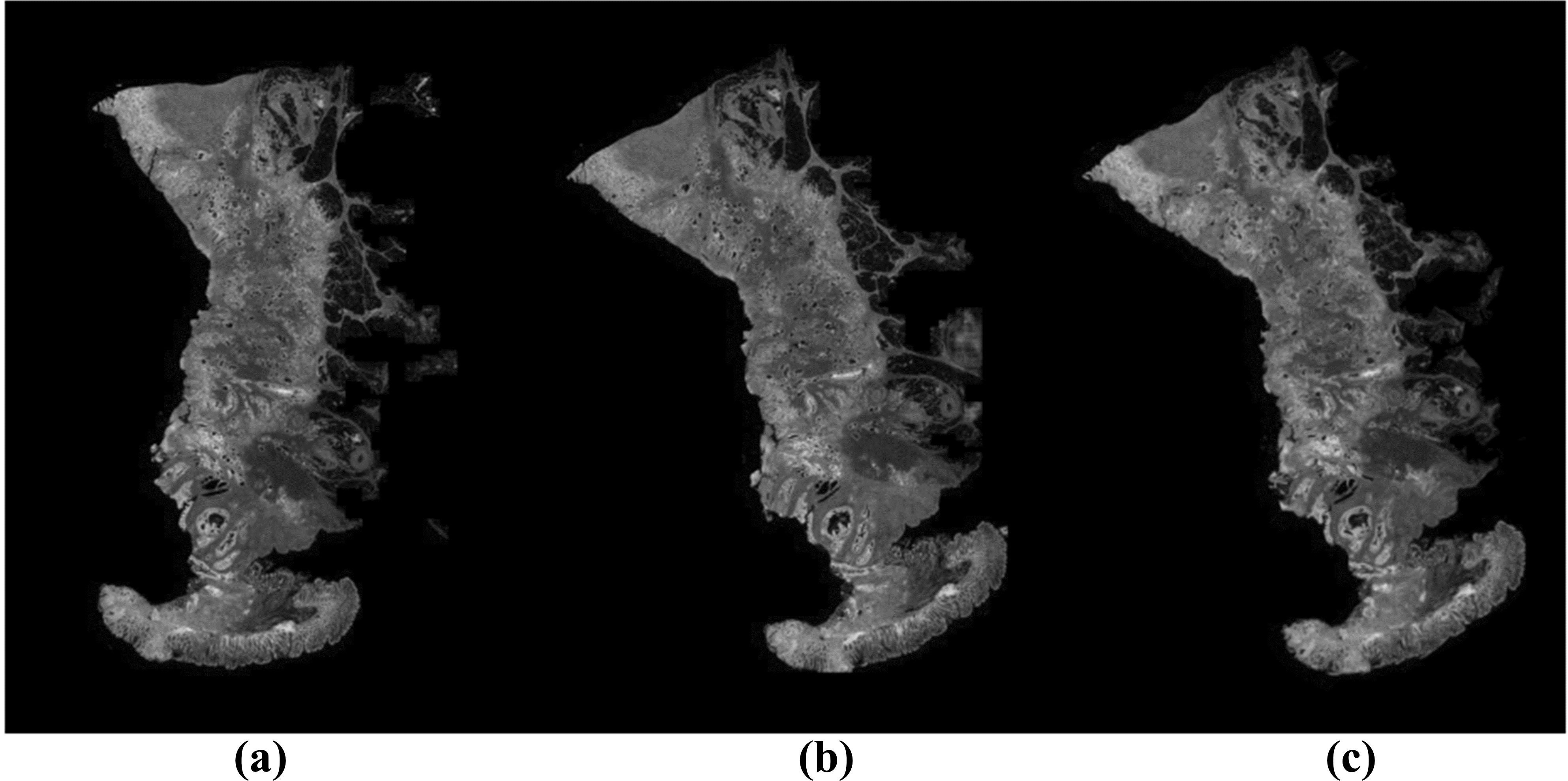}
\caption{An example of the alignment of a pair of images on the ANHIR dataset is provided in Fig, where (a) (b) (c) are the source image, the target image, and the transformed source image, respectively. After the deformation length is applied to the source image, we can observe the comparison with the target image.}
\label{fig:result0}
\end{figure}

\begin{figure*}
\centering
\includegraphics[scale=0.4]{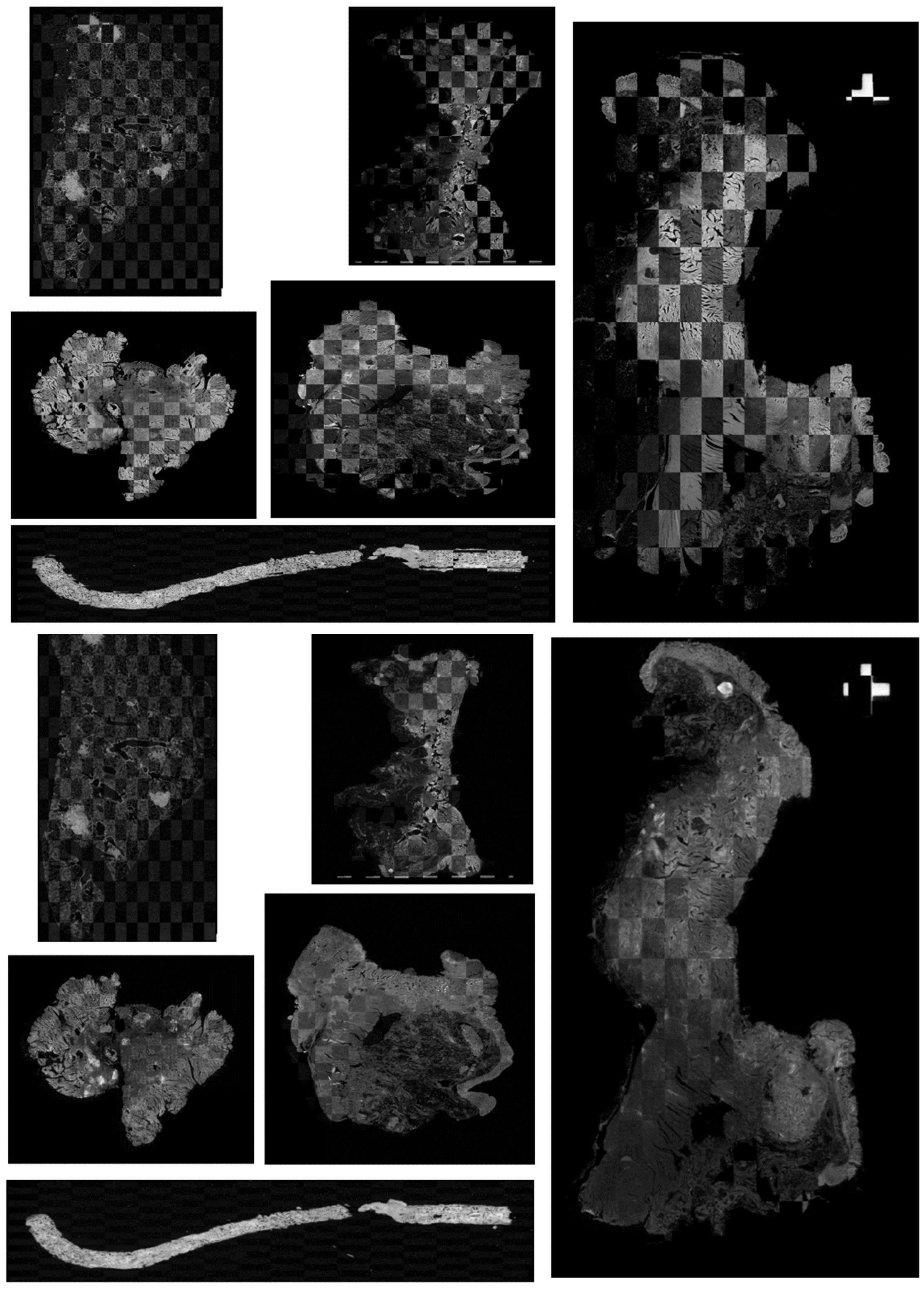}
\caption{The figure shows the results of comparing the target image with the registered image, and the results are presented in the form of the checkerboard. The target image and the aligned image alternately appear in the figure. The registration results for the general contour and texture are excellent. The distinction between the source image and the target image is given in the first figure. The comparison between the transformed source image and the target image is illustrated in the second figure.}
\label{fig:figure_9}
\end{figure*}

\subsection{Fine-Tune Moudle}

We use the final matches obtained above and the paired landmarks given by the training set to construct a new training set based on the structural features of the query points matching Network.

\begin{table*}
\centering
\caption{Quantitative results evaluated on the ANHIR dataset}
\resizebox{\textwidth}{!}{%
\begin{tabular}{llllllllll}
\hline
 &
  Average-Average rTRE &
  Average-Median rTRE &
  Median-Average rTRE &
  Median-Median rTRE &
  Max-Average rTRE &
  Max-Median rTRE  \\ \hline
\multicolumn{10}{c}{ALL}                                                                      \\ \hline
Ours                 & \textbf{0.0035} &\textbf{0.0022} & \textbf{0.0025}  & \textbf{0.0016}          & 0.0258  & 0.0170 \\
Detector-based only      & 0.0056 & 0.0031 &0.0032  & 0.0021  & 0.0397  & 0.0225 \\
Detector-free only            & 0.0059  &  0.0028 &0.0035 & 0.0019 & 0.0546 & 0.0245 \\
SFG*   \cite{b111}        & 0.0056  & 0.0024 & 0.0045 & \textbf{0.0016} & 0.0240 & \textbf{0.0156} \\
MEVIS\cite{b7}             & 0.0043  & 0.0028 & 0.0028 & 0.0018 & 0.0251 & 0.0188 \\
AGH\cite{b24}            & 0.0073  &0.0032  &0.0036  & 0.0017 & 0.0290 & 0.0214 \\
UPENN\cite{b25}             & 0.0041  & 0.0029 & 0.0029 & 0.0019 & \textbf{0.0238} & 0.0224 \\
DeepHistReg*\cite{b4}             & 0.0060  &0.0033  &0.0047  & 0.0019 & 0.0239 & 0.0224 \\
CKVST\cite{b1}             & 0.0042  & 0.0026 & 0.0027 & 0.0023 & 0.0239 & 0.0189 \\
TUNI\cite{b1}           & 0.0063  & 0.0031 & 0.0048 & 0.0021 & 0.0287 & 0.0204 \\
 \\ \hline
\multicolumn{10}{c}{TEST}                                                                 \\ \hline
Ours                 & \textbf{0.0037} &\textbf{0.0023}  &\textbf{0.0026}  & \textbf{0.0017}          & 0.0277  &\textbf{0.0183}  \\
SFG*          & 0.0082  &0.0026  &0.0070  & \textbf{0.0017} & 0.0286 &\textbf{0.0183}  \\
MEVIS            & 0.0044  & 0.0027  &0.0029 & 0.0018 & 0.0251 & 0.0188 \\
AGH           & 0.0072  & 0.0031 & 0.0057 &\textbf{0.0017}  & 0.0251 & 0.0188 \\
UPENN            & 0.0041  & 0.0029 & 0.0029 & 0.0019 & \textbf{0.0239} & 0.0190 \\
DeepHistReg*    & 0.0061 &  0.0030  &0.0047 & 0.0019 & 0.0276 & 0.0197 \\
CKVST           & 0.0042  & 0.0026 & 0.0027 & 0.0023 & 0.0239 & 0.0189 \\
TUNI          & 0.0063  & 0.0031 & 0.0048 & 0.0021 & 0.0287 & 0.02045 & & & \\ \\ \hline
\end{tabular}%
}
\label{table1}
\end{table*}

\begin{table*}
\centering
\caption{Quantitative results evaluated on the ANHIR dataset after fine-tuned}
\resizebox{\textwidth}{!}{%
\begin{tabular}{llllllllll}
\hline
 &
  Average-c &
  Average-Median rTRE &
  Median-Average rTRE &
  Median-Median rTRE &
  Max-Average rTRE &
  Max-Median rTRE  \\ \hline
\multicolumn{10}{c}{ALL}                                                                      \\ \hline
Ours*(train)  & \textbf{0.0020} & \textbf{0.0012} & \textbf{0.0011} & 0.0008         & \textbf{0.0130}  & \textbf{0.0047} \\
Ours                 & 0.0035 & 0.0022 & 0.0025 & 0.0016          & 0.0258  & 0.0170 \\
Detector-based only      & 0.0056 & 0.0031  &0.0032 & 0.0021  & 0.0397  & 0.0225 \\
Detector-free only            & 0.0059  & 0.0028 &0.0035  & 0.0019 & 0.0546 & 0.0245 \\
SFG*(supervised)    &0.0046   &\textbf{0.0010}  &0.0038  & \textbf{0.0007} &  0.0186 & 0.0073 \\
SFG*           & 0.0081  &0.0024  &0.0069  & 0.0016 & 0.0284 & 0.0172 \\
MEVIS            & 0.0043  & 0.0028 & 0.0028 & 0.0018 & 0.0251 & 0.0188 \\
AGH           & 0.0073  & 0.0032 & 0.0036 & 0.0017 & 0.029 & 0.0214 \\
UPENN            & 0.0041  & 0.0029 & 0.0029 & 0.0019 & 0.0238 & 0.0224 \\
DeepHistReg*            & 0.0060  & 0.0033 & 0.0047 & 0.0019 & 0.0239 & 0.0224 \\
CKVST           & 0.0042  & 0.0026 &0.0027  & 0.0023 & 0.0239 & 0.0189 \\
TUNI          & 0.0063  & 0.0031 &0.0048  & 0.0021 & 0.0287 & 0.0204 \\
 \\ \hline
\multicolumn{10}{c}{TEST}                                                                 \\ \hline
Ours*(train) & \textbf{0.0034} & \textbf{0.0022} & \textbf{0.0023} & \textbf{0.0016}          & \textbf{0.0240}  &\textbf{0.0169}  \\

Ours                 & 0.0037 & 0.0023 & 0.0026 & 0.0017         & 0.0277  &0.0183  \\
SFG*(supervised)  & 0.0083  &0.0025  &0.0070  & \textbf{0.0016} & 0.0291 &0.0181  \\
SFG*          & 0.0082  &0.0026  &0.0070  & 0.0017 & 0.0286 &0.0183  \\
MEVIS            & 0.0044  &0.0027  &0.0029  & 0.0018 & 0.0251 & 0.0188 \\
AGH           & 0.0072  &0.0031  &0.0057  & 0.0017 & 0.0251 & 0.0188 \\
UPENN            & 0.0041  & 0.0029 & 0.0029 & 0.0019 & \textbf{0.0239} & 0.0190 \\
DeepHistReg*    & 0.0061 & 0.0030  &0.0047  & 0.0019 & 0.0276 & 0.0197 \\
CKVST           & 0.0042  &  0.0026 &0.0027 & 0.0023 & 0.0239 & 0.0189 \\
TUNI          & 0.0063  & 0.0031 &0.0048  & 0.0021 & 0.0287 & 0.02045 & & & \\ \\ \hline
\end{tabular}%
}
\label{tabel2}
\end{table*}

Since the input image of the detector-free network is only 256 × 256, we use the iterative method to perform inference in multi-scale images, as shown in Fig. \ref{fig:trainset}. We down-sample the image after getting the previous inference result and crop the image with the inference point as the center to get a new image for iterative inference. This operation enables the algorithm to obtain accurate matching results on large-scale images. We use both the ground truth given in the dataset and the final match obtained in the previous inference process as the dataset. Since the final results obtained in the previous model are more accurate and robust than those obtained by a single detector-free network, the pseudo-labels enhance the robustness of the model here. We constructed different levels of paired images, as shown in Fig. \ref{fig:trainset}. Training is performed using pseudo-labels and real labels together.

The new loss function is

\begin{footnotesize} 
 	\begin{equation}
 		\begin{aligned}
 			\underset{\boldsymbol{\Phi}}{\arg \min } \underset{\left(\boldsymbol{x}, \boldsymbol{x}^{\prime}, \boldsymbol{I}, \boldsymbol{I}^{\prime}\right) \sim \mathcal{D}}{\mathbb{E}} (\mathcal{L}_{\text {corr }}+\mathcal{L}_{\text {cycle }})_{Pseudo-Label} +(\mathcal{L}_{\text {corr }}+\mathcal{L}_{\text {cycle}})_{Label}
 		\end{aligned}
 	\end{equation}
 \end{footnotesize}

After fine-tuning, the model has been significantly improved. In the fine-tune process, we use both ground-truth given by the training set of ANHIR and the pseudo-label we obtained as the training set. The original backbone is frozen and the learning rate of the transformer is 1e-6.

\section{Results}
The source and target images are fed into the two networks. The final matched point pairs are retrieved after anomaly matching detection. We interpolate the collected matches to obtain the dense DVF. Finally, the DVF is applied to the source image to obtain the transformed image, depicted in Fig. \ref{fig:result0}.

\begin{figure}
\centerline{\includegraphics[scale=0.32]{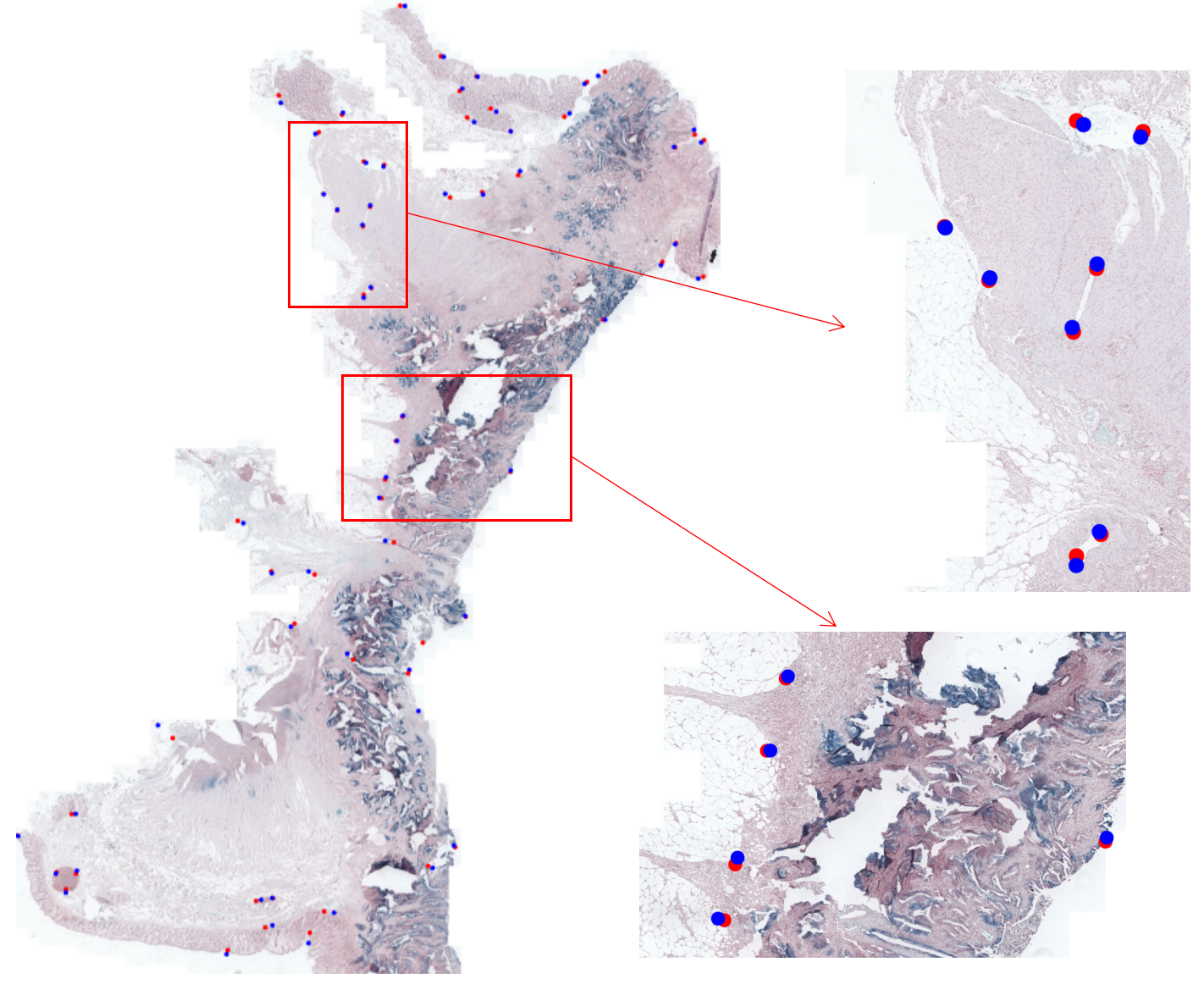}}
\caption{The alignment results are shown in the form of landmarks. The red dots are the results obtained by model inference, and the blue dots are ground-truth.}
\label{fig:present}
\end{figure}

We compared the feature points obtained from a network based on deep learning features with traditional features. The evaluation was carried out on various datasets. It was observed that the deep learning-based approach had improved significantly compared to the traditional approach regarding the number of point pairs and reliability. Fig. \ref{fig:compare} depicts the impact of the comparison.

Fig. \ref{fig:compare} displays the matching results of four distinct tissue pathology images using various point matching methods. The two rows above are the matching results of the traditional methods, SIFT and ORB, respectively. While the two rows below are the matching results of the two deep learning networks employed in this paper. It is apparent that the points derived from the two rows of images above are quite sparse and have a significant mismatch rate. Inversely, the matched point pairs generated using the deep learning approach cover the whole image and every part of the organization, allowing us to utilize these accurate and dense matched point pairs for interpolation to approximate the real DVF.

Fig. \ref{fig:figure_9} exhibits a checkerboard representation of our registration result. As demonstrated in the figure, the contours and internal textures of the registered image and the target image are extremely closely matched, and the junction of the checkerboard is quite smooth. We randomly selected a graph to compare the inference results with the landmark results in Fig. \ref{fig:present} We found that our prediction of landmarks is highly accurate. We evaluate our results on the grand challenge website. Table \ref{table1} summarizes the test results as compared to other techniques. We only submitted the results from the pre-trained model without any targeted training. We found that our method has a significant advantage over other methods in terms of the median and the mean of rTRE. We also provide the outcomes just with matched network interpolation. According to the table, utilizing a single matching network has proven extremely competitive. Our total accuracy surpasses any other technique. The results in all data sets and the test set were provided. The method marked with * in the table is a deep learning-based method that was trained in the training set. The rest are conventional methods.

After fine-tuning, our results have been further improved, as shown in Table \ref{tabel2}. The diagram below summarizes our results. And our results are outstanding in practically every measure and achieved the state-of-the-art performance in the Average rTRE on the test dataset (\href{https://anhir.grand-challenge.org/evaluation/challenge/leaderboard/}{\underline{Challenge Leaderboard}}). The rTRE (relative Target Registration Error) metric is calculated as follows:

\begin{footnotesize} 
 	\begin{equation}
 		\begin{aligned}
            r T R E=\frac{T R E}{\sqrt{w^{2}+h^{2}}}
 		\end{aligned}
 	\end{equation}
 \end{footnotesize}

Where $w$ and $h$ are the image dimensions, TRE is the MSE of the predicted and true values.

\begin{figure}
\centerline{\includegraphics[scale=0.6]{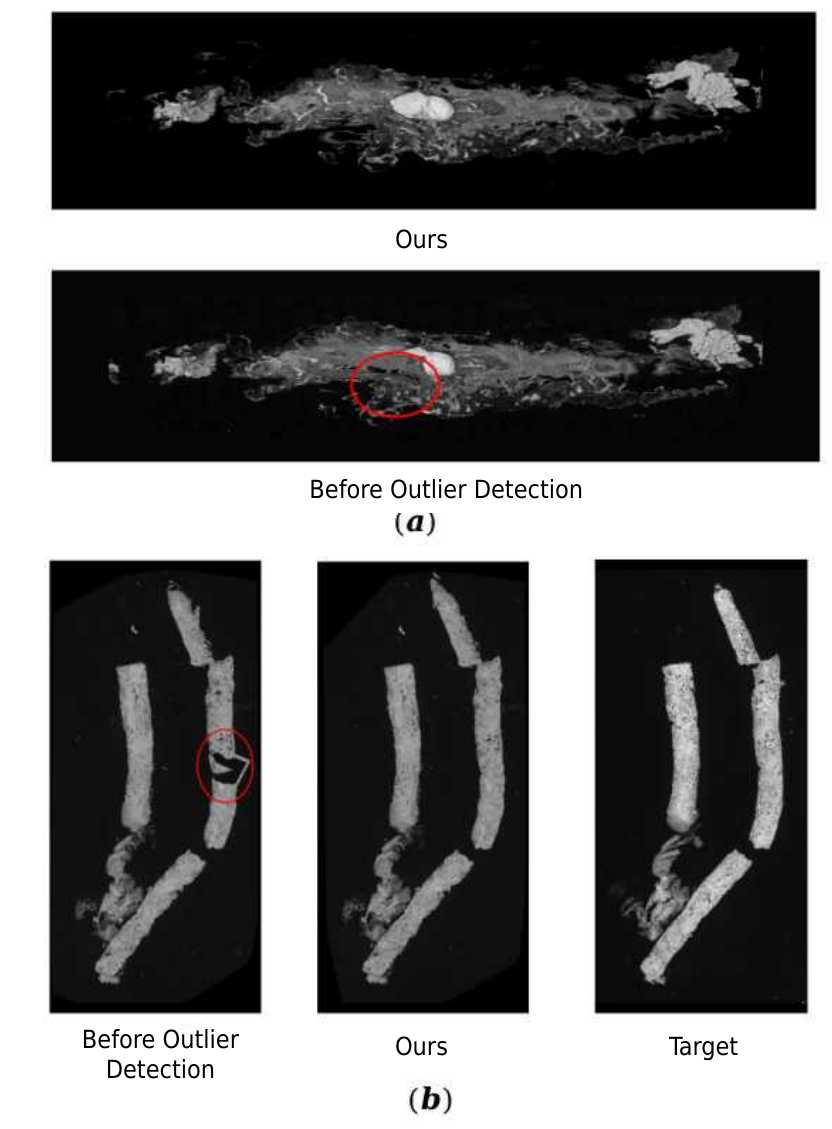}}
\caption{Figures (a) and (b) show the registration results using direct interpolation without outlier detection module. Compared with the results from the proposed framework, our method solves the image distortion problem caused by mismatch.}
\label{fig:compare_new}
\end{figure}

We compare the results without outlier detection module and after using the outlier detection module to filter out the mismatching pairs in Fig. \ref{fig:compare_new}. Our modified model greatly improves the accuracy of the feature point matching model. It can be observed from the part in the red circle that the image without outlier detection has severe distortion, which heavily impacts the final accuracy and the availability of the registration network. After the outlier detection eliminates the false matches, the image distortion passing through the DVF vanishes and becomes flatter. It indicates that our statistical anomaly matching detection model is effective.

\section{Discussion}
The experimental results validate that our proposed hybrid feature-based deep learning registration framework effectively addresses the challenges of traditional feature matching. Our approach employs two deep learning feature networks that extract a denser set of feature points. Moreover, the proposed outlier detection model effectively eliminates false matches. The effectiveness of these methods is quantitatively validated on the ANHIR dataset.

Building upon this success, we employed, for the first time, the approach of fine-tuning natural image pre-training models in medical image data for pathology image registration. This innovative approach proved effective in the experiment, demonstrating that natural image pre-training model fine-tuning holds significant potential for application in medical images. This success can be attributed to the large dataset on natural images that enables powerful generalization, raising the question of whether it is possible to compensate for the lack of comprehensive data in medical images by using a large number of natural pictures.

Interestingly, our studies revealed that we could achieve relatively robust performance using merely the pre-trained models from natural images without targeted training. This may be due to the large dataset on natural images that enables powerful generalization. However, several drawbacks exist in the proposed approach.

First, we extract feature points and match them via a hybrid network, resulting in a more complex network structure. Consequently, our approach significantly underperforms conventional iteration-based algorithms in terms of computing efficiency. Second, regularization terms and differential analysis may restrict the DVF's smoothness, preserving the image's topology in deformation-based neural networks and conventional iteration-based methods. In contrast, if there is a mismatch, the feature point matching-based framework may disregard the image's structure.

To address these limitations, our future work will focus on improvement in two directions: (1) to design more lightweight networks that maintain accuracy while reducing computational complexity; (2) to apply this framework to other medical images, such as computed tomography (CT) and magnetic resonance imaging (MRI). We note that registration has a broad range of applications for the 3D medical image scenario, including assisted segmentation, tumor tracking, target area outlining, etc. Consequently, we plan to extend our deep feature registration network to 3D, contributing to enhanced accuracy in large deformation and multi-modal registration.

In addition to these improvements, further exploration could involve integrating different types of features, such as texture, shape, and intensity, to enhance the robustness of the registration process. By combining various features, the proposed method could potentially achieve superior performance in challenging situations, such as images with significant noise or deformation. Furthermore, incorporating spatial information into the feature extraction process could also improve the performance of the proposed method, particularly in cases where the images have a high degree of deformation or noise.

Another promising direction for future work is the development of unsupervised or semi-supervised learning techniques for feature extraction and matching. This could potentially help overcome the limitations of relying on labeled data, especially in the medical imaging domain, where obtaining accurate annotations can be challenging and time-consuming. By leveraging unsupervised or semi-supervised learning, the proposed method could adapt to new and diverse datasets more efficiently, reducing the need for extensive manual annotation.

Expanding on this idea, the proposed method could be extended to handle multi-modal image registration, where the images being registered are acquired using different imaging modalities (e.g., CT and MRI). This is a common problem in medical imaging and presents additional challenges, such as differences in image intensity, contrast, and resolution. By addressing these challenges, the proposed method could become an even more versatile and valuable tool for medical image registration.

Lastly, to ensure the robustness of our approach, evaluating the proposed method on larger and more diverse datasets would provide further insights into its generalizability and robustness. This could include comparisons with state-of-the-art registration methods and the investigation of different network architectures and training strategies to optimize performance. By rigorously evaluating the proposed method on a wide range of datasets, we can better understand its strengths and limitations, as well as identify areas for further refinement.

\section{Conclusion}

This paper proposes a novel registration network based on hybrid deep features. The proposed framework outperforms the current state-of-the-art methods in terms of accuracy. We particularly overcome the two feature point challenges of pair sparsity and mismatching as opposed to standard feature-based techniques. Moreover, our network achieves excellent results just by pre-training on natural images. After fine-tuning it, we get better results, which indicates that our approach is more generalizable and may be utilized for image registration in clinical and applications in medicine.

\section*{Acknowledgments}
This work is partly supported by grants from the National Natural Science Foundation of China (82202954, U20A201795, U21A20480, 12126608) and the Chinese Academy of Sciences Special Research Assistant Grant Program.

\section*{Declarations}

 The authors declare that they have no known competing financial interests or personal relationships that could have appeared to influence the work reported in this paper.

\end{document}